\documentclass[aps,prb,reprint,twocolumn,groupedaddress,floatfix,showpacs]{revtex4-1}
\usepackage[usenames,dvipsnames]{pstricks}
\usepackage{epsfig,graphics,graphicx}
\usepackage{amsmath,amsfonts,amssymb,latexsym}
\usepackage{dsfont}

\begin{document}

\title{Quantum and classical correlations in antiferromagnetic  chains \\ and the realization of Werner states with spins}

{\frenchspacing

\author{P. R. Wells Jr.}  \email{wells@if.ufrj.br}

\affiliation{Instituto de F{\'\i}sica, Universidade Federal do Rio de Janeiro Cx.P. 68528, 21945-970, RJ, Brazil}

\author{Belita Koiller}

\affiliation{Instituto de F{\'\i}sica, Universidade Federal do Rio de Janeiro Cx.P. 68528, 21945-970, RJ, Brazil}

\date{\today}

\begin{abstract}

We investigate pairwise correlation properties of the ground state (GS) of finite antiferromagnetic (AFM) spin chains described by the Heisenberg model. The exchange coupling is restricted to nearest neighbor spins, and is constant $J_0$ except for a pair of neighboring sites, where the coupling $J_1$ may vary.  We identify a rich variety of possible behaviors for different measures of pairwise (quantum and classical) correlations and entanglement in the GS of such spin chain. Varying a single coupling affects the degree of correlation between all spin pairs, indicating possible control over such correlations by tuning $J_1$. We also show that a class of two spin states constitutes exact spin realizations of Werner states (WS). Apart from the basic and theoretical aspects, this opens concrete alternatives for experimentally probing non-classical correlations in condensed matter systems, as well as for experimental realizations of a WS via a single tunable exchange coupling in a AFM chain.

\end{abstract}


\maketitle
}

\section{Introduction}

Quantum behavior at the microscopic scale is well established both theoretically and experimentally. Strictly quantum phenomena observed so far have confirmed the validity of quantum mechanics postulates,  in particular its non-local character.\cite{aspect1981} Entanglement is certainly one of the most intriguing  and basic of such quantum phenomena, as it is a fundamental resource in performing most tasks in quantum information and quantum computation. \cite{nielsen}  
Recently it has been realized that entanglement is not the only form of quantum correlations, and that separable quantum states,  that were expected to show strictly classical behavior, in fact can exhibit quantum correlations other than entanglement. A measure of quantum correlations that has recently received much attention is the quantum discord. \cite{PhysRevLett.88.017901} It has been analyzed both theoretically \cite{jphys-a34-6899-hend-vedral, PhysRevA.77.042303, sarandy:022108, PhysRevA.81.042105, ferraro2010almost, PhysRevLett.105.190502, PhysRevA.82.032112} and experimentally. \cite{soares2010, PhysRevLett.107.070501, PhysRevLett.107.140403} 

Experiments demonstrating and quantifying entanglement have mostly dealt with photons \cite{walborn2006, almeida2007environment, jimenez2009determining} and atoms. \cite{anderson1995observation, yu2009sudden}
Solid state systems could in principle provide more compact and stable hosts to study such effects. It is also desirable to demonstrate control over pair-related quantum phenomena in solid state systems, such as spins of electrons bound to quantum dots \cite{PhysRevA.57.120, PhysRevB.59.2070} or to impurities  \cite{kane1998} in semiconductors.

We investigate here pairwise correlations and entanglement in small (few sites) magnetic chains. The magnetic moments are the spins of electrons individually bound to a short linear array of  quantum dots in GaAs or Si-based nanostructures, or of donors in Si. Neighboring spins interact via exchange coupling. An important aspect of such fabricated chains is the tunability \cite{footnote1} of the coupling between spins in specified pairs. \cite{Petta30092005, PhysRevLett.107.030506, Nowack02092011} Changes in this coupling affect the pair correlations, and we identify general trends to be expected when one coupling varies in such chains. As a general trend obtained, the behavior of the different correlations studied is similar, seeming to illustrate a \textit{monogamous} nature \cite{PhysRevA.61.052306, PhysRevLett.96.220503, PhysRevA.78.012325}, even for correlations other than entanglement. Also, we show that a class of two-spin states of even-numbered chains are exact realizations of Werner states (WS). These states are useful as they simulate the effect of the environment in destroying coherence and entanglement in a quantum system.

This paper is organized as follows. Our model is presented in Sec.~\ref{sec:model}, where the Hamiltonian and the density matrix are introduced. Measures quantifying correlations and entanglement used here are briefly reviewed in Sec.~\ref{sec:theor-bg}. In Sec.~\ref{sec:results} the behavior of small model chains is presented. In Sec.~\ref{sec:werner} we obtain and discuss the realization of WS in our spin model.  Our final conclusions are given in Sec.~\ref{sec:concl}.

\section{\label{sec:model} Model Hamiltonian and two-spin density matrices}

We consider the isotropic Heisenberg model for antiferromagnetic (AF) spin-half finite chains. The Hamiltonian for this system is given by
\begin{equation}
H = \sum_{\langle i,j \rangle}^{N} J_{ij} \, \mathbf{S}_i \cdot \mathbf{S}_j \, ,
\label{eq:heisenberg}
\end{equation}
where the summation is over first neighbors of an open chain with $N$ spins, and the exchange coupling is positive ($J_{ij} > 0$). We obtain the spectrum for the Hamiltonian in Eq.~(\ref{eq:heisenberg}) in the product basis of the $S_i^z$ eigenstates, of the form $| s_1 s_2 \cdots s_N \rangle \equiv | s_1 \rangle \otimes | s_2 \rangle \otimes \cdots \otimes | s_N \rangle $ where $s_i$ represents up or down spin at site $i$, i.e., $ \vert s_i \rangle =\{ \vert \! \! \uparrow \rangle_i, \vert \! \! \downarrow \rangle_i \}$. The total spin operator in the $z$-direction, $S_{z}^{T} = \sum_{i=1}^{N} S_{i}^{z}$, is a good quantum number, as $\left[ H, S_{z}^{T} \right] = 0$, and so the Hamiltonian exhibits a block-structure, each block having a well-defined value of $S_{z}^{T}$. For AF coupling, the lowest state is in the lowest $S_{z}^{T}$ subspace, so even-$N$ chains have  a non-degenerate $S_{z}^{T} = 0$ ground state (GS). Odd-$N$ chains GS is doubly-degenerate, one state from  each block $S_{z}^{T} = 1/2$ or $S_{z}^{T} = -1/2$.

The simplest AF system is that of two spins, for which the GS is a singlet:
\begin{equation}
| S \rangle = \frac{1}{\sqrt{2}} \left( \vert \! \! \uparrow \downarrow \rangle - \vert \! \! \downarrow \uparrow \rangle \right) \, \, .
\label{eq:singlet}
\end{equation}
This is a maximally entangled state, in the sense that knowledge of the global state of the two spins gives no
information about the state of either spin, while by choosing an observable, measurement of the state of either spin gives with certainty the state of the other. This quantum correlation is stronger than could ever be possible classically.

The two-spin density matrices $\rho_{ij}$ for spins located at sites $i$ and $j$, are obtained from the full
density matrix $\rho_0 = \vert \psi_0 \rangle \langle \psi_0 \vert$ by tracing out the degrees of freedom of the other
$N-2$ spins. These are in general mixed states, written as a weighted sum of pure states:
$\rho_{ij} = \sum_{\ell} p_\ell \rho_\ell$, where $p_{\ell}$ is the weight of $\rho_\ell=  \vert \psi_\ell \rangle \langle \psi_\ell \vert$.

It has been shown that the symmetries of the Hamiltonian (\ref{eq:heisenberg}) imply that the two-spin density matrix
for any pair of spins in the $S_z$ basis $\{ | \! \! \uparrow \uparrow \rangle, | \! \! \uparrow \downarrow \rangle, | \! \! \downarrow \uparrow \rangle, | \! \! \downarrow \downarrow \rangle \}$, has the general form \cite{sarandy:022108, PhysRevA.69.022311}
\begin{equation}
\rho_{ij} = \left(
\begin{array}{cccc}
 a & 0 & 0 & 0 \\
 0 & b_1 & z & 0 \\
 0 &  z& b_2 & 0\\
 0 & 0 & 0 & d
\end{array} \right) \, \, ,
\label{eq:densitymatrix}
\end{equation}
with $\text{Tr} \rho_{ij} = 1$, $b_1 b_2 \ge |z|^2$ and $a d \ge 0$. \cite{PhysRevA.81.042105} The single spin
density matrix $\rho_i$ corresponding to spin $i$ is obtained by tracing out spin $j$.

\section{\label{sec:theor-bg} Quantifying correlations}

Several quantities describe correlations in quantum systems; we summarize below those adopted in our study.
The amount of information contained in a quantum system, assumed here to be a spin array described by $\rho$, is traditionally measured through the Von Neumann entropy,
\begin{equation}
{\cal{S}} (\rho) = - \, \text{Tr} \, \rho \log_{2} \rho \, \, ,
\label{eq:vnentropy}
\end{equation}
which is the quantum generalization of the classical Shannon entropy. When spins $i$ and $j$ interact, they usually share some information, and this shared information may be quantified by the quantum mutual information (QMI),
\begin{equation}
{\cal{I}} (\rho_{ij}) = {\cal{S}}(\rho_i) + {\cal{S}}(\rho_j) - {\cal{S}}(\rho_{ij}) \, \, ,
\label{eq:qmi-v1}
\end{equation}
which includes all (quantum and classical) correlations between spins $i$ and $j$. QMI may also be defined as
\begin{equation}
{\cal{J}}(\rho_{ij}) = {\cal{S}}(\rho_i) - {\cal{S}}(\rho_{i|j}) \, \, ,
\label{eq:qmi-v2}
\end{equation}
where $S (\rho_{i|j})$ is the quantum conditional entropy, i.e., the entropy of spin $i$ when the state of spin $j
$ is known. Conceptually, obtaining $S(\rho_{i|j})$ involves measurement over one spin, say $j$. Restricting ourselves to projective measurements, denoted by the set of projectors $\{ \Pi_{m}^{j} \}$, each possible state of the two-spin system after the measurement is of the form
\begin{equation}
\rho_m = \frac{1}{p_m} \big( \mathds{1}_i \otimes \Pi_{m}^{j} \big) \rho_{ij} \big( \mathds{1}_i \otimes \Pi_{m}^{j} \big) \, \, ,
\end{equation}
where
\begin{equation}
p_m = \text{Tr} \big( \mathds{1}_i \otimes \Pi_{m}^{j} \big) \rho_{ij} \big( \mathds{1}_i \otimes \Pi_{m}^{j} \big)
\end{equation}
is the probability of outcome $m$ and $\mathds{1}_i$ is the identity operator for spin $i$. Quantum conditional entropy is then defined as a weighted sum of the entropies associated to the possible states $\rho_m$,
\begin{equation}
{\cal{S}}(\rho_{i|j}) = \sum_m p_m \, {\cal{S}}(\rho_m) \, \, .
\end{equation}
Classically,  ${\cal{I}}$ and ${\cal{J}}$ are equivalent, but the quantum generalizations may differ.  Their difference defines Quantum Discord, \cite{PhysRevLett.88.017901}
\begin{equation}
\delta(\rho_{ij}) = {\cal{I}}(\rho_{ij}) - {\cal{J}}(\rho_{ij}) \, \, ,
\label{eq:qdiscord}
\end{equation}
which also depends on the measurement basis. Note that, contrary to $\cal I$,  $\cal J$  depends on the measurement basis used to specify the state of spin $j$ and is not symmetric with respect to interchange $i\leftrightarrow j$.
Quantum correlation $(Q^C)$ is defined as
\begin{equation}
Q^C (\rho_{ij}) = \min_{ \{ \Pi_{m}^{j} \} } \delta(\rho_{ij}) \, \, ,
\label{eq:qcorr}
\end{equation}
where the minimization is with respect to all possible measurement basis sets. This minimization amounts to
finding the measurement basis that minimally disturbs the system, in particular extracting
information about spin $j$ with minimum disturbance on spin $i$. Defining classical correlation $(C^C)$ as
\cite{jphys-a34-6899-hend-vedral, PhysRevA.77.042303, sarandy:022108}
\begin{equation}
C^C (\rho_{ij}) = \max_{ \{ \Pi_{m}^{j} \}} {\cal{J}}(\rho_{ij}) \, \, ,
\label{eq:ccorr}
\end{equation}
it is possible to decompose the  total correlation as a sum of quantum and classical contributions:
\begin{equation}
{\cal{I}}(\rho_{ij}) = Q^C (\rho_{ij}) + C^C (\rho_{ij}) \, \, .
\end{equation}
The individual components $Q^C$ and $C^C$ are not
symmetric in $i\leftrightarrow j$, since ${\cal{J}}$ is not, while their sum ${\cal{I}}$ is symmetric.

Finally, the degree of entanglement between two spins is also investigated here through the usual measure,  the
concurrence,  \cite{PhysRevLett.78.5022, PhysRevLett.80.2245}
\begin{equation}
{\cal{C}} = \max \{ 0, \Lambda \} \, \, ,
\label{eq:conc}
\end{equation}
with
\begin{equation}
\Lambda = \sqrt{\lambda_1} - \sqrt{\lambda_2} - \sqrt{\lambda_3} - \sqrt{\lambda_4} \, \, .
\end{equation}
The four numbers $\lambda_{i = 1, \cdots ~ 4}$ are the eigenvalues in decreasing order of the operator $\rho_{ij}
\tilde{\rho}_{ij}$, with
\begin{equation}
\tilde{\rho}_{ij} = ( \sigma_y \otimes \sigma_y ) \rho_{ij} ( \sigma_y \otimes \sigma_y ) \, \, ,
\end{equation}
where $\sigma_y$ is the second Pauli matrix. Concurrence has the property $0 \le {\cal{C}} \le 1$, with ${\cal{C}} = 0$ indicating that the state is separable while ${\cal{C}} = 1$ for a maximally entangled state, such as a two-spin singlet.
Regarding spins interchange, we note that ${\cal C}$ is symmetric: ${\cal C}_{ij} = {\cal C}_{ji}$.

The concurrence for systems described by the density matrix of Eq.~(\ref{eq:densitymatrix}) can be directly obtained from Eq.~(\ref{eq:conc}) and is given by
\begin{equation}
{\cal{C}} = 2 \max ( 0, |z| - \sqrt{ad} ) \, \, .
\label{eq:conc-analytic}
\end{equation}
Taking for example the singlet state of Eq.~(\ref{eq:singlet}), we have ${\cal{I}} (= 2) = C^C (= 1)+ Q^C( = 1)$ and $
{\cal{C}} = 1$. These values are extreme (maxima), meaning that the singlet is maximally correlated, in the classical and in the quantum sense.

In what follows, when there is no ambiguity, we refer in general to all quantities ${\cal C}$, $I$, $C^C$ and $Q^C$ as \textit{correlations}.

\section{\label{sec:results} General trends}

We analyze here the behavior of correlations for two-spin states of even- and odd-$N$ linear chains as a function of the exchange coupling parameter $\xi \equiv J_{23}/J_{12} = J_{1} / J_{0}$. Explicitly we treat in this section the cases of $N = 3$ and $N = 4$, as depicted, respectively, in Figs.~\ref{fig:3sc}(a) and \ref{fig:3sc}(b). In these chains, the coupling $J_1$ varies, while the others are kept constant and equal to $J_0$. For $N = 4$ we consider also the more general case where $J_{34} = J_2 \neq J_0$. Exact diagonalization of these chains were obtained here both analytically and numerically as a function of $\xi$.

\subsection{Three-spins chain}

%-------figure-----interacting spins chain in the first scheme here------------
\begin{figure}[b!]
\includegraphics[width=.9\columnwidth]{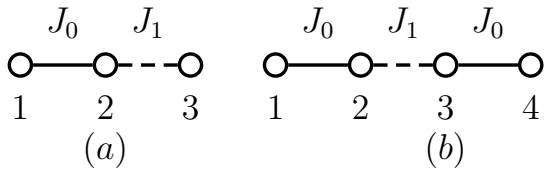}
\caption{\label{fig:3sc} Simple spin configurations with (a) three and (b) four spins, with a variable coupling
between spins $2$ and $3$. The other couplings are fixed to $J_0$.}
\end{figure}
%--------------------------------------
%
\begin{figure}[b!]
\includegraphics[width=\columnwidth]{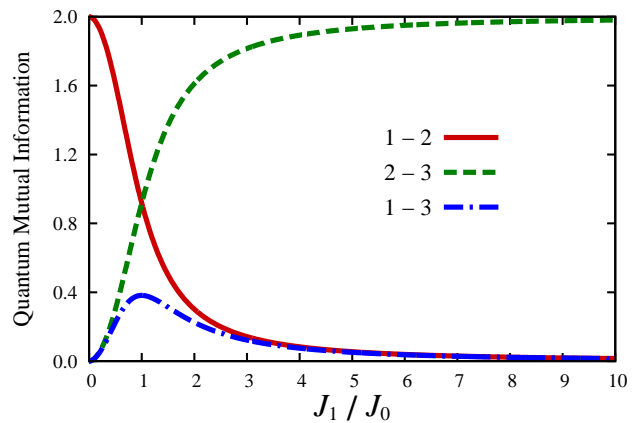}
\caption{\label{fig:imq3spins-escheme1} (Color online) QMI as a function of $\xi = J_1/J_0$ for the three-spins system of Fig.~\ref{fig:3sc}(a). The full, dashed and dashed-dotted lines refer respectively to the pairs $1 - 2$, $2 - 3$ and $1 - 3$.}
\end{figure}
\begin{figure}[b!]
\includegraphics[width=\columnwidth]{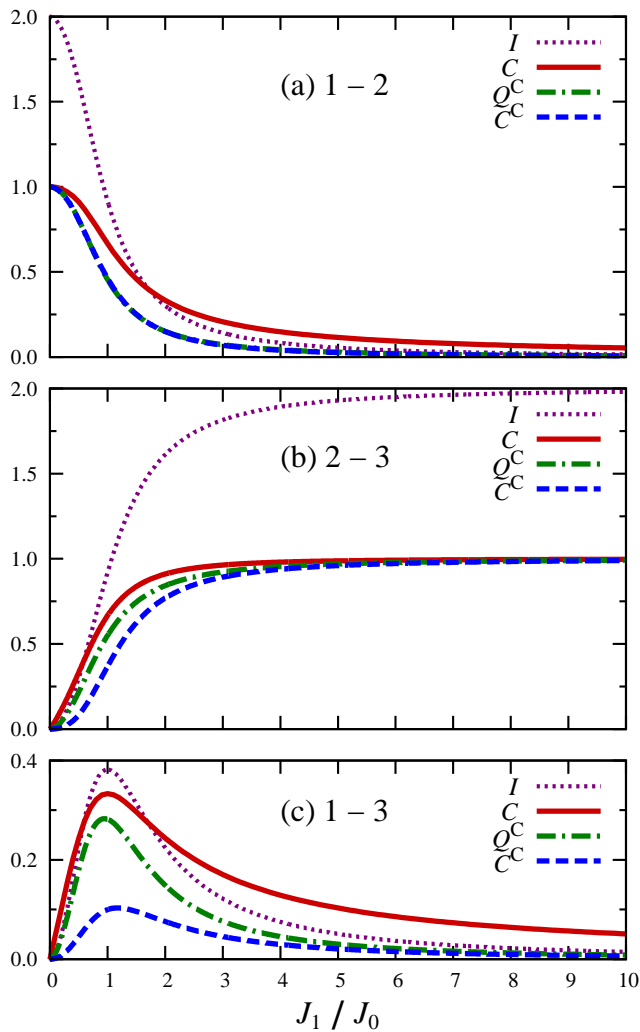}
\caption{\label{fig:corr+conc-3spins} (Color online) Total, classical and quantum correlations and concurrence as a
function of $\xi  = J_1/J_0$, for the three-spins system at Fig.~\ref{fig:3sc}(a). For comparison, we show (dotted line)
the same data as in Fig.~\ref{fig:imq3spins-escheme1} for ${\cal{I}}$. {\cal{C}}, $Q^C$ and $C^C$ are plotted respectively in full, dash-dotted and dashed lines, for pairs (a) 1 - 2, (b) 2 - 3 and (c) 1 - 3. See text for details.}
\end{figure}
As the GS of a three-spins chain is twofold degenerate, we need to choose one of the subspaces, say that of $S_z^T = 1/2$ (considering units in which $\hbar = 1$). The two degenerate states can be separated by an applied magnetic field.
The basis for the $S_z^T = 1/2$ subspace reads as $\{ | \! \! \uparrow \uparrow \downarrow \rangle, | \! \! \uparrow \downarrow \uparrow \rangle, | \! \! \downarrow \uparrow \uparrow \rangle \}$, and the general form of the GS state vector is given by
\begin{equation}
| \psi_0^{(3)} \rangle = \alpha_1 | \!\! \uparrow \uparrow \downarrow \rangle + \alpha_2 | \! \uparrow \downarrow \uparrow \rangle + \alpha_3 | \! \downarrow \uparrow \uparrow \rangle \, \, .
\label{eq:3spinsGSvector}
\end{equation}
Above, $\alpha_2$ is the amplitude for the classical N\'eel state, and $\alpha_1$ and $\alpha_3$ are the amplitudes of the quantum fluctuations. For $J_1 = J_0$ we obtain $\alpha_1 = \alpha_3 = - 1 / \sqrt{6}$ and $\alpha_2 = 2 / \sqrt{6}$. By explicit calculation it is possible to obtain the two-spin reduced density matrix for the state vector above. We have $d = 0$ at Eq.~\ref{eq:densitymatrix}, for two spins pointing down is not allowed in this subspace [for $S_z^T = - 1/2$, $a = 0$ in Eq.~(\ref{eq:densitymatrix})]. The general form of the density matrix may be written as
\begin{equation}
\rho_{ij} = p_1 \, | \! \uparrow \uparrow \rangle \langle \uparrow \uparrow \! | + p_2 | \phi \rangle \langle \phi | \, \, ,
\label{eq:densitymatrix-3spins}
\end{equation}
where
\begin{equation}
| \phi \rangle = c_1 | \! \uparrow \downarrow \rangle + c_2 | \! \downarrow \uparrow \rangle
\end{equation}
is a general singlet, with $|c_1|^2 + |c_2|^2 = 1$. This state has the property that it is maximally correlated only when $|c_1| = |c_2|$. This implies that the two-spin state in the chain is a mixture of an entangled component $| \phi \rangle \langle \phi |$ with a separable one, $| \! \! \uparrow \uparrow \rangle \langle \uparrow \uparrow \! \! |$, the relative weights given by $p_1$ and $p_2$.
Fig.~\ref{fig:imq3spins-escheme1} shows ${\cal{I}}$ as a function of $\xi$ for the configuration of Fig.~\ref{fig:3sc}(a). As $J_{1}$ increases, ${\cal{I}}_{12}$ decreases, while ${\cal{I}}_{23}$ increases; they cross at $J_1 = J_0$ as, by symmetry, pairs $1-2$ and $2-3$ become equivalent at this point. As regards to the second-neighbor spins $1$ and $3$, they also get correlated, although only nearest neighbors are exchanged-coupled. This is an effective interaction, mediated by the central spin $2$, as discussed at Ref.~\onlinecite{PhysRevB.82.140403} within the context of \textit{long distance entanglement}. For $J_1 \ll J_0$, behavior of pair $1 - 3$ follows that of the pair $2 - 3$, and for $J_1 \gg J_0$ it approaches that of pair $1 - 2$. Also, ${\cal{I}}_{13}$ exhibits a maximum at $J_1 = J_0$, corresponding to the tendency of these two spins of forming an entangled triplet state $(| \! \! \uparrow \downarrow \rangle + | \! \! \downarrow \uparrow \rangle)/\sqrt{2}$, as can be verified by explicitly constructing $\rho_{13}$ at $J_1 = J_0$.
The corresponding $C^C$, $Q^C$ and ${\cal C}$, are presented and compared to ${\cal{I}}$ in Fig.~\ref{fig:corr+conc-3spins}, where we see that the overall behavior of these four correlations is qualitatively the same. These results show that correlations tend to concentrate in the most strongly exchange-coupled pair, as explained by the following argument. Two isolated spins get maximally correlated for any finite value of the AFM exchange coupling. When they are interacting with additional spins, all neighboring pairs will tend to form a singlet, but this is formally not possible.  The most strongly coupled pair forms a state with a larger singlet character than all others, thus becoming the most correlated, although not maximally. \cite{PhysRevB.82.140403} It becomes clear from these results that whenever two spins are maximally correlated, this precludes any correlation with another third spin, so that in general, as long as two spins share some amount of correlation, their correlation with a third spin is limited. This result is related to the so-called \textit{monogamous nature of entanglement}, \cite{PhysRevA.61.052306, PhysRevLett.96.220503, PhysRevA.78.012325} and the present results seem to indicate that it applies to all other (quantum and classical) correlations.

Correlations in this three-spins chain exhibit properties associated with the symmetry of the system which are easily verified by inspection of the spin arrangement and are also obtained algebrically.  Namely $\rho_{12} (\xi) = \rho_{32} (1 / \xi)$, $\rho_{21} (\xi) = \rho_{23} (1 / \xi)$, $\rho_{13} (\xi) = \rho_{31} (1 / \xi)$. All symmetries of $\rho$ are readily transfered to all correlation functions described here. We recall that ${\cal C }$ and ${\cal{I}}$ are symmetric with respect to spins interchange, while $\rho$, $Q^C$ and $C^C$ in general are not. Finally, we remark that in all cases we have $C^{C} \leq Q^{C} \leq {\cal{C}}$.

\subsection{\label{sub4} Four-spins chain}

We perform a systematic study of pair correlations and entanglement for the particular $N=4$ chain in Fig.~\ref{fig:3sc}(b), where $J_{12}=J_{34}=J_0$ and the variable coupling is $J_{23}=J_1$. This keeps the same number of free parameters as the N=3 case. We discuss the more general situation  $J_{12}\ne J_{34}$ at the end of this section.

Different from the $N = 3$ case, here we have spin interchange symmetry for $\rho$   ($\rho_{ij}=\rho_{ji}$) and for all correlation functions. It is possible to show by explicitly writing the full GS density matrix on the $S_{z}^{T} = 0$ four-spins basis, that the reduced density matrices $\rho_{12} = \rho_{34}$, $\rho_{13} = \rho_{24}$ and $\rho_{14} = \rho_{23}$ for all $\xi$. As a consequence,  all correlations follow the same symmetry properties. Regarding the $\xi \leftrightarrow 1 / \xi$ correspondence given above due to the symmetry of the $N = 3$ chains, here we find algebrically a modified variable transformation, namely $\rho_{12} (\xi) = \rho_{23} (4 / \xi) $ which is not obvious to anticipate from symmetry arguments and, again, is transferred to all correlations.

\begin{figure}[t!]
\includegraphics[width=\columnwidth]{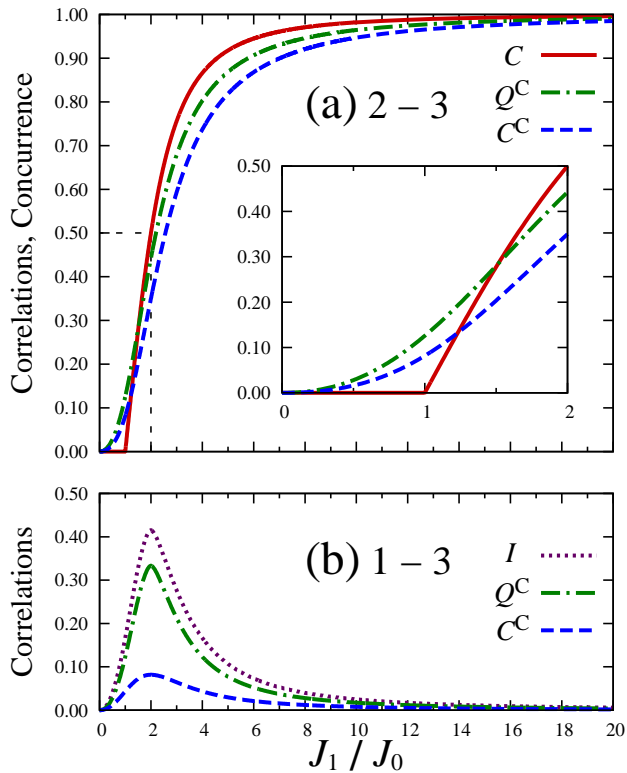}
\caption{\label{fig:4spins} (Color online) Correlations as a function of $\xi  = J_1/J_0$ for selected spin pairs of the GS of the four-spins chain depicted in Fig.~\ref{fig:3sc}(b). (a) The full, dash-dotted and dashed lines are respectively for ${\cal{C}}$, $Q^C$ and $C^C$ of pair $2-3$. The inset shows the region $0 \leq \xi \leq 2$. (b) Dotted, dash-dotted and dashed lines refer to total ($I$), quantum ($Q^C$) and classical ($C^C$) correlations for the pair $1-3$, whose concurrence is zero in the full range. See text for details.}
\end{figure}

Fig.~\ref{fig:4spins} shows correlations as functions of $\xi$ for pairs (a) $2-3$ and (b) $1-3$. As in the $N = 3$ case (Figs.~\ref{fig:corr+conc-3spins}(b) and \ref{fig:corr+conc-3spins}(c) respectively), the overall behaviors of ${\cal C}$, $Q^C$ and $C^C$ are similar, and each pair shows $C^C(\xi)<Q^C(\xi)$ for all $\xi$. However, here an entanglement/disentanglement transition is obtained at special values of $\xi$, while for $N = 3$ the pairs entanglement vanishes only at limiting situations. We have $C_{23} = 0$ for $0 \leq \xi < 1$, and for $\xi > 1$ the pair becomes entangled. From this point on, $C_{23}$ increases rapidly, reaches the value $1/2$ at $\xi = 2$, and asymptotically tends to unity. This limit corresponds to a singlet formed by spins $2$ and $3$. Independently, although in parallel, spins $1$ and $4$ also tend to form a singlet, as anticipated by the property $\rho_{23} (\xi) = \rho_{14} (\xi)$. The other property mentioned, vis., the correspondence $\xi \leftrightarrow 4 / \xi$ between pairs $1-2$ and $2-3$ permits one to infer the behavior of $C_{12}$ from that of $C_{23}$: $C_{12}$ is unit at $\xi = 0$, and decreases to $1/2$ at $\xi = 2$. It vanishes for $\xi \geq 4$.

As regards to the other correlations, $Q^{C}_{12} \, ( = Q^{C}_{34})$ is unit for $\xi = 0$, and decreases gradually, tending to zero asymptotically for large $\xi$, and $Q^{C}_{23} \, ( = Q^{C}_{14})$ increases from zero at $\xi = 0$ and asymptotically tends to one. In both cases $Q^{C}$ remains non-zero in regions in which concurrence vanishes, indicating that separable states present quantum correlations not captured by concurrence. A similar description holds for $C^{C}_{12} \, (= C^{C}_{34})$ and $C^{C}_{23} (= C^{C}_{14})$ [See Fig.~\ref{fig:4spins}(a)]. Here we also have $C^{C} \leq Q^{C}$, but ${\cal{C}}$ can be less or greater than $C^{C}$ and $Q^{C}$. Fig.~\ref{fig:4spins}(b) shows that total, classical and quantum correlations are nonzero for the pairs $1-3$ and $2-4$ and exhibit a maximum at $\xi = 2$, while ${\cal{C}}_{13} = {\cal{C}}_{24} = 0$ over the full range.

In order to test the robustness and generality of our results, we consider the case in which $J_{34} = J_2 \neq J_0$. This is a likely experimental situation due to disorder. We find that there is always an entanglement/disentanglement transition, but at different values of $\xi = J_1/J_0$, according to the value of $J_2$. Our numerical results are consistent with the following transition points $\xi_C$ for the pairs $1-2$ (entangled for $\xi < \xi^{12}_{C}$) and $2-3$ (entangled for $\xi > \xi^{23}_{C}$):
\begin{eqnarray}
\xi_C^{12} &=& 2 (1 + J_{34}/J_{12}) \, \, ,  \\
\xi_C^{23} &=& (1 + J_{34}/J_{12})/2 \, \, .
\label{eq:xic-23}
\end{eqnarray}
Using the values $J_{12} = J_{34} = J_0$ we obtain $\xi^{23}_C = 1$ [as indicated in Fig.~\ref{fig:4spins}], and $J_{12} = J_{34} = J_0$, $\xi^{12}_{C} = 4$.

We remark that symmetries discussed above, considering $J_{12} = J_{34} = J_0$, remain valid in the general case where $J_{34} = J_2 \neq J_0$.

\section{\label{sec:werner} Werner states}

As noted in Sec.~\ref{sub4}, an entangled/disentangled transition may occur for specific pairs for the $N=4$ chain at finite values of $\xi$, e.g, ${\cal C}_{23}$ in Fig.~\ref{fig:4spins}(a) vanishes for $0<\xi<1$ but not for $\xi>1$.
As this behavior resembles that of the so-called Werner states (WS),\cite{werner1989quantum}
we investigate here two-spins states in even-numbered chains to establish if and how they relate to WS. A WS is a one-parameter state that represents a mixture of a maximally entangled state with a totally mixed component proportional to the identity matrix. \cite{PhysRevA.66.062315,jin2010quantum, PhysRevA.70.022321} We consider here the following description, suited to the case at hand:
\begin{equation}
\rho_W = \frac{1 - p}{4} \, \mathds{1}_4 + p \, | \Psi_{-} \rangle \langle \Psi_{-} | \, \, ,
\label{eq:werner-def}
\end{equation}
where  $\mathds{1}_4$ is the identity matrix in the four-dimensional space, $| \Psi_{-} \rangle = (| 0 1 \rangle - | 1 0 \rangle )/\sqrt{2}$ is one of the Bell states,\cite{PhysRevA.66.062315} analogous to the singlet state of Eq.~(\ref{eq:singlet}), and the parameter $p$ ($0 \le p \le 1$) gives  the weight of the entangled component, driving an  entanglement $(p>1/3)$/disentanglement $(p<1/3)$ transition.

The GS of even-$N$ AF chains considered here is a $S_z^T = 0$ state, for which the two-spins density matrix elements $\rho_{ij}$ in Eq.~(\ref{eq:densitymatrix}) may be written in terms of a single parameter,\cite{PhysRevA.69.022311} namely
\begin{equation}
a = d = \frac{1}{4} + \Gamma_{ij} \, , \, b_1 = b_2 = \frac{1}{4} - \Gamma_{ij} \, , \, z = 2 \Gamma_{ij} \, \, ,
\label{eq:coefs-n-par}
\end{equation}
where
\begin{equation}
\Gamma_{ij} = \langle S_i^z S_j^z \rangle \, \,
\label{eq:corrfunc}
\end{equation}
is the spin-spin correlation function. The relations (\ref{eq:coefs-n-par}) and (\ref{eq:corrfunc}) above are direct implications of rotational and time reversal symmetries of the Hamiltonian. Eq.~(\ref{eq:densitymatrix}) implies
\begin{equation}
\rho_{ij}^{N \rm even} = \frac{ 1 + 4 \Gamma }{4} \, \mathds{1}_4 - 4 \Gamma \, | S \rangle \langle S | \, \, ,
\end{equation}
where $| S \rangle$ is the singlet state from Eq.~(\ref{eq:singlet}). From the definition of $\cal{C}$ in Eq.~(\ref{eq:conc-analytic}) and from the bounds $-1/4 \le \Gamma \le 1/4$, it follows  that when $ \Gamma > -1/12$ the concurrence vanishes and the state is separable. For $\Gamma < 0$ the density matrix has the same structure given by Eq.~(\ref{eq:werner-def}), with the weight parameter $p = - 4 \Gamma$, preserving the bounds $0 \le p \le 1$. Thus, the density matrix $\rho_{ij}$ of a spin pair of an even-$N$ AF chain GS is a genuine WS when $\Gamma < 0$. Given that $\Gamma$ is a function of the coupling ratio  alone,
$\Gamma = f (\xi)$,
we may conclude that, in principle, controlling the coupling $J_1$ allows control over the fundamental parameter that characterizes a WS.

\subsection{\label{sub34}Three and four-spins chains}

\begin{figure}[b!]
\includegraphics[width=.95\columnwidth]{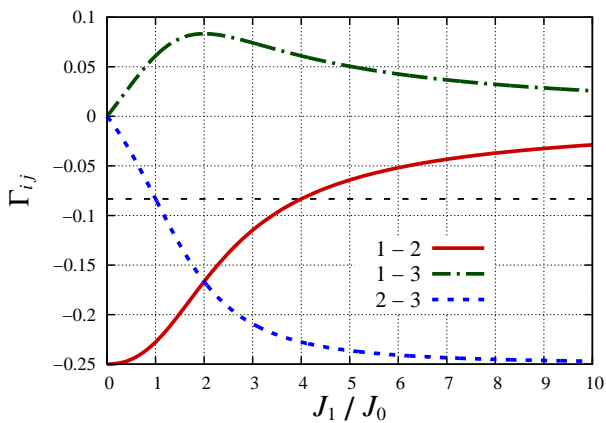}
\caption{\label{fig:gammaxi} (Color online) Spin-spin correlation function, $\Gamma_{ij}$, as a function of the coupling ratio for the four-spins chain in Fig~\ref{fig:3sc}(b). We find $\Gamma_{13} > 0$, ${\cal{C}}_{13} = 0$, in agreement with Fig.~\ref{fig:4spins}(b). The horizontal dashed line, given by $\Gamma = -1/12$, crosses $\Gamma_{12}$ and $\Gamma_{23}$, respectively, at the transition points $\xi^{12}_{C} = 4$ and $\xi^{23}_{C} = 1$.}
\end{figure}

As an illustration of the criteria above, we show in Fig.~\ref{fig:gammaxi} the spin-spin correlation functions for the four-spins chain of Fig~\ref{fig:3sc}(b). The correlations are always negative for antiferromagnetically coupled pairs, namely $\Gamma_{12} (= \Gamma_{34})<0$ and $\Gamma_{23} (= \Gamma_{14})<0$, so these pairs constitute representations of  WS.
On the other hand, $\Gamma_{13} (= \Gamma_{24})>0$, representing a general separable state $({\cal C}=0)$. This means that, for the $S_z^T = 0$ states considered, the effective coupling between second-neighbors spins is ferromagnetic, leading to positive $\Gamma$ and null entanglement. This condition is not valid  in general for odd-$N$ chains ($S_z^T = \pm 1/2$ states), as for example the three-spins chain where although the spins $1-3$ are coupled via a ferromagnetic effective interaction ($\Gamma_{13}>0$), they are (weakly) entangled (${\cal C}_{13} \ne 0)$ for any $\xi$.
\begin{figure}[]
\includegraphics[width=\columnwidth]{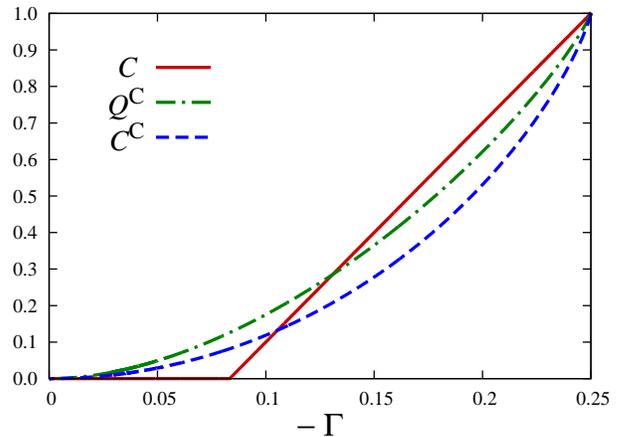}
\caption{\label{fig:corrxgamma} (Color online)  Correlations and concurrence as a function of minus the spin-spin correlation function in the $\Gamma < 0$ region for a four-spins chain. The curves are universal for pairs $1-2$, $2-3$, $3-4$, and $1-4$.}
\end{figure}
\begin{figure}[]
\includegraphics[width=\columnwidth]{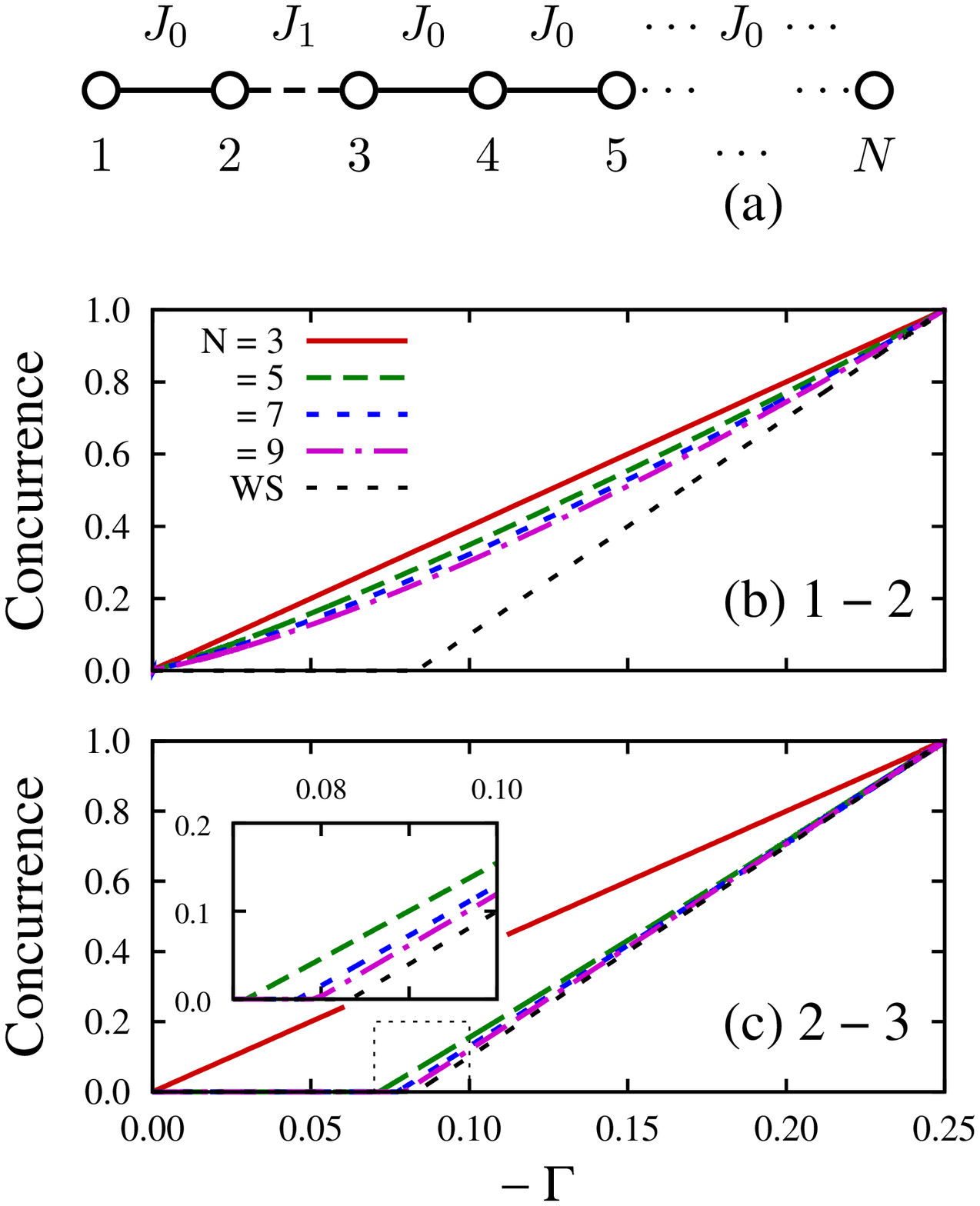}
\caption{\label{fig:corr-oddNxgamma} (Color online) (a) Schematic representation of spin chains considered in the paper. The ``surface'' spin is labeled ``1'' and all couplings between nearest neighbors equal $J_0$, except for $J_{23} = J_1$.  (b) and (c) show concurrence as a function of the negative of the spin-spin
correlation function, $\Gamma$, for odd-$N$ chains as represented in (a). (b) Results for the $1 - 2$ pair for $N = 3$, $5$, $7$ and $9$. (c) Results for the $2 - 3$ pair; the inset shows the rectangular dotted region. For comparison, the result for the WS is also shown in Figs.~\ref{fig:corr-oddNxgamma}(b) and (c).}
\end{figure}

When the relation $\Gamma = f(\xi)$ is invertible for a pair $i-j$, as is the case for $\Gamma_{12}$ and $\Gamma_{23}$ for $N=4$ (See Fig.~\ref{fig:gammaxi}), correlations for  $i-j$ may be explicitly obtained as functions of $\Gamma$. This leads to the result that, for pairs with $\Gamma < 0$, correlations dependence on $\Gamma$ is universal. For example, $C_{12}$ and $C_{23}$ [Fig.~\ref{fig:4spins}(a)] collapse into\cite{PhysRevA.69.022311}
\begin{equation}
{\cal{C}} = 6 \max \{ 0, -\Gamma - 1/12 \} \, \, , \, \, \Gamma < 0 \, \, ,
\label{eq:concws}
\end{equation}
given by the full line of Fig.~\ref{fig:corrxgamma}. The quantum and classical correlations for these pairs,  represented in Fig.~\ref{fig:corrxgamma} are also universal: $Q^{C}_{12}$ and $Q^{C}_{23}$ [Fig.~\ref{fig:4spins}(a)] collapse into the dash-dotted line of Fig.~\ref{fig:corrxgamma}, while $C^{C}_{12}$ and $C^{C}_{23}$ into the dashed line. We remark that for the more general case where $J_{34} = J_2 \neq J_0$, the two-spins states analyzed are still exact realizations of WS. This is a very interesting point in an experimental point of view, for it may happen to be difficult to make $J_{12}$ exactly equal to $J_{34}$.

Regarding the $N = 3$ chain, the pair density matrix in Eq.~(\ref{eq:densitymatrix}) for the GS shows  either $a = 0$ or $d = 0$ (for $S_z^T = -1/2$ or $S_z^T = 1/2$ respectively).  Eq.~(\ref{eq:conc-analytic}) now gives  ${\cal{C}} = 2 |z| = 4 |\Gamma|$ for all pairs. We show ${\cal{C}}_{12}$ and ${\cal{C}}_{23}$ for $N = 3$ in Fig.~\ref{fig:corr-oddNxgamma}(a) and (b), respectively.

\subsection{\label{subgeneral} Longer chains}

For completeness, we consider spin chains extending the ones in Fig.~\ref{fig:3sc} following the scheme of Fig.~\ref{fig:corr-oddNxgamma}(a), where spin 1 remains at the ``surface''.
We have verified that for longer ($N>4$) even-numbered chains, the pairs $1-2$ and $2-3$ are still exact representations of WS. Other pairs may or may not exhibit an entanglement/disentanglement transition, only with a different functional form for ${\cal C}(\Gamma)$ due to the interactions of spin 4 with additional spins (as compared with the $N=4$ chain where 1 and 4 are equivalent by symmetry).

For odd-numbered chains, ${\cal{C}}_{12}$ and ${\cal{C}}_{23}$ for $N = 3,5,7$ and $9$ are presented in  Figs.~\ref{fig:corr-oddNxgamma}(b) and \ref{fig:corr-oddNxgamma}(c). It is clear that the results are qualitatively different for the two selected spin pairs.  The behavior of ${\cal{C}}_{12}$ is remnant of the linear dependence on $|\Gamma|$ exhibited by the $N = 3$ chains [Fig.~\ref{fig:corr-oddNxgamma}(b)] while ${\cal{C}}_{23}$ for odd $N \ge 5$ presents a entanglement/disentanglement transition at a value of $\Gamma$ that approaches the one for a WS ($-1/12$) for increasing $N$, as shown in Fig.~\ref{fig:corr-oddNxgamma}(c). We infer that zero total spin, $S_z^T=0$, is not a requirement to obtain entanglement/disentanglement transitions, even though the involved states may not be mapped into the standard WS.

\section{\label{sec:concl} Discussions and Conclusions}

Our study of quantum and classical pair correlations in the GS of open linear AF spin chains devotes particular attention to short chains, namely $N=3$ and $N=4$ chains. The qualitatively distinct behavior of even and odd$-N$ chains, already well established, \cite{{PhysRev.135.A640, *PhysRevLett.90.047901, *PhysRevB.68.134417, *chaves:104410, *chaves:032505}} is further illustrated by correlation properties.

For a class of states in even-numbered chains (here fully exemplified by $N = 4$), we identify two-spin density matrices as exact representations of WS, and we present the  relationship between the WS weight parameter and the spin-spin correlation function of the spin pair. In our $N = 4$ example, we considered a chain with a tunable central bond $J_1$ while the others are fixed to $J_0$. We remark that if the ``fixed'' bonds are different, i.e., if the bonds sequence is $J_0 - J_1 - J_2$, the entanglement/disentanglement transition takes place at a different value of $J_1$, which is now a function of $J_2$. We verified that even in this more general case exact WS representations are obtained. These results point to a possible experimental WS realization in the framework of solid state systems, namely the GS of an AFM spin chain. This is particularly interesting because so far WS realizations were mainly proposed within quantum optics frameworks. \cite{PhysRevA.66.062315,jin2010quantum, PhysRevA.70.022321} We note that our result was obtained for zero temperature. It has been shown that the WS mixing coefficient may be associated with the temperature of a one-dimensional Heisenberg spin model. \cite{Batle200512}

Finally we suggest possible physical systems for experimental verification of WS behavior in spin chains. The candidates include: (i) donor-bound electron spins in a donors array precisely positioned in Si. Such arrays with up to four P donors are within fabrication capabilities, as discussed in Ref.~\onlinecite{doi:10.1021/nl2025079} (ii) Quantum dots arrays with single electron occupation each. Similar four-dots arrays fabrication were also recently reported \cite{PhysRevLett.107.030506}; (iii) magnetic atoms nanochains. \cite{hirjibehedin2006spin}. In systems (i) and (ii), control over $J_1$ is possible through nano-electrodes aligned to the respective bond. Observation of WS in simple condensed matter systems as proposed here seems to be accessible with current nano-processing capabilities.

\begin{acknowledgments}

We acknowledge J. d'Albuquerque e Castro and Thereza Paiva for valuable suggestions and  A. L. Saraiva and C. M. Chaves for interesting conversations.
This work was partially supported by the Brazilian agencies CNPq, CAPES and FAPERJ, under the programs PAPDRJ-2010, Cientista do Nosso Estado and Pensa-Rio.

\end{acknowledgments}

\bibliographystyle{apsrev4-1}
%\bibliography{refs}

\begin{thebibliography}{10}
\bibitem{footnote1}%
  \BibitemOpen
  \bibinfo {note} {By varying gate voltages that raise or lower the
  electrostatic barrier between the electrons in the pair.}%
  \bibAnnoteFile{Stop}{footnote1}%
\end{thebibliography}

%Merlin.mbs v4.21 2009-07-09.
%

\end{document}